\documentclass[12pt,preprint]{aastex}

\newcommand{\etal}{et al.}

\shorttitle{CO Line Emission from Compact Nuclear Starburst Disks Around AGN}
\shortauthors{Armour and Ballantyne}

\begin{document}

\title{CO Line Emission from Compact Nuclear Starburst Disks Around Active Galactic Nuclei}


\author{J. N. Armour\altaffilmark{1} and D. R. Ballantyne\altaffilmark{1}}
\altaffiltext{1}{Center for Relativistic Astrophysics, School of Physics, Georgia
  Institute of Technology, 837 State Street, Atlanta, GA 30332-0430;
  jarmour3@gatech.edu}

\begin{abstract}
There is substantial evidence for a connection between star formation
in the nuclear region of a galaxy and growth of the central
supermassive black hole.  Furthermore, starburst activity in the
region around an active galactic nucleus (AGN) may provide the
obscuration required by the unified model of AGN. Molecular line
emission is one of the best observational avenues to detect
and characterize dense, star-forming gas in galactic nuclei over a range of redshift. This paper presents predictions for the carbon
monoxide (CO) line features from models of nuclear starburst disks around AGN.
These small scale ($\la 100$~pc), dense and hot starbursts have CO
luminosities similar to scaled-down ultra-luminous infrared
galaxies and quasar host galaxies. Nuclear starburst disks that
exhibit a pc-scale starburst and could potentially act as the
obscuring torus show more efficient CO
excitation and higher brightness temperature ratios than those without
such a compact starburst. In addition, the compact starburst models predict
strong absorption when $J_{\mathrm{Upper}} \ga 10$, a unique observational signature of these objects. These findings allow for the possibility
that CO SLEDs could be used to determine if starburst disks are
responsible for the obscuration in $z \la 1$
AGN. Directly isolating the nuclear CO line emission of such compact
regions around AGN from galactic-scale emission will require high
resolution imaging or selecting AGN
host galaxies with weak galactic-scale star formation. Stacking
individual CO SLEDs will also be useful in detecting the
predicted high-$J$ features. 
\end{abstract}


\section{Introduction}
\label{sect:intro}
The growth and structure of active galactic nuclei (AGN) and their
effect on the development of their host galaxies are amongst the most
popular and intensive areas of research in modern astrophysics.  These
objects are believed to consist of a central supermassive black hole
(SMBH) surrounded by an accretion disk (e.g., Lynden-Bell 1969; Shakura
\& Sunyaev 1973).  The infalling material of the disk produces the
characteristic emission of the AGN.  Despite this accepted universal
mechanism to explain the appearance of AGN, the properties of the
emission of a particular nucleus can vary greatly and are believed to
be explained by the unified model of AGN, which accounts for the
differences between various sub-classes of active galactic nuclei by
positing the existence of extensive obscuration in the region near the
AGN (e.g., Antonucci \& Miller 1985; Maiolino \& Rieke 1995).  This
obscuration is required to be anisotropic by the available evidence
and is believed to be caused by material concentrated in a torus
around the active galactic nucleus (Antonucci 1993).

However, multiple observations have now shown that this simple unified picture cannot
hold over all redshifts and AGN luminosities. The obscured fraction of
AGNs decreases with luminosity \citep[e.g.,][]{ueda03,laf05,simp05,ak06,has08,tkd08,tuel08,db11} and seems to increase with
redshift \citep[e.g.,][]{bem06,tu06,has08}. Well-sampled monitoring of some local Seyfert galaxies show
rapid changes of X-ray obscuring columns, indicating a dynamic and
clumpy medium \citep[e.g.,][]{ris10}. Global population synthesis models that include
both X-ray and optical/near-IR constraints indicate that the unified
model seems to hold only for $z \la 1$ \citep{db11}. These results all point
to a scenario where there are different origins for the AGN
obscuration that depend on the redshift and luminosity of the central
engine \citep[e.g.,][]{bem06,has08}.

A possible origin for the obscuring torus in $z < 1$ Seyfert galaxies is a nuclear starburst
disk (e.g., Wada \& Norman 2002; Ballantyne 2008; Wada et al. 2009)
--- compact ($\la 100$~pc) regions of strong star-formation that,
through a combination of radiation pressure and supernova feedback,
can potentially inflate an optically thick structure and obscure the
central AGNs. Indeed, simple 1-D analytic models (Ballantyne 2008) showed an
AGN could be both obscured and fueled by a pc-scale ultra-compact starburst embedded
in a larger nuclear starburst disk. Due to the competition for gas
between star-formation and accretion, this model could only obscure
Seyfert galaxies, but these are just the AGNs that dominate the hard
X-ray background \citep[e.g.,][]{ueda03,laf05,db10}. Thus, it is useful to pursue the possible
observational signatures of the nuclear starburst model that can be
tested with the large multiwavelength AGN samples produced by the
deep X-ray surveys \citep[e.g.,][]{xue11}. To that end, Ballantyne (2008) discussed
strategies to detect these starburst disks at mid-infrared and
radio wavelengths, leading Pierce et al. (2011) to uncover evidence
for the predicted levels of star-formation in radio-stacks of $z<1$
X-ray selected AGNs.

The greatest difficulty in observationally testing the nuclear
starburst disk model is separating out its emission from the AGN and
the host galaxy. The high angular resolution allowed by radio
interferometric observations therefore seems to be the most promising
technique for further investigation. With the ongoing construction of
the Atacama Large Millimeter/Submillimeter Array (ALMA), capable of reaching an angular resolution of 5 milliarsec at 650 GHz,
it is therefore interesting to consider the molecular line properties
of the nuclear starburst disk models. The region within several dozen parsecs of an AGN and its central
supermassive black hole is known to contain a wealth of dense and
relatively warm gas (e.g. Scoville et al. 1991; Hseih et al. 2008;
Wada et al. 2009; Papadopoulos et al. 2010a).  These conditions of
relatively high density ($n_{H_{2}}\geq 10^{3-4}$ cm$^{-3}$) and high
pressure favor the creation of significant quantities of molecular gas
(e.g. Pelupessy et al. 2006). It is well-known that star
formation occurs solely in molecular gas (e.g. Fukui \& Kawamura 2010;
Schruba et al. 2011). For reasons outlined in a variety
of sources (e.g. Dickman et al. 1986; Fukui \& Kawamura 2010), carbon
monoxide (CO) is the molecule most often used to probe regions of
molecular gas.  Despite the importance of CO lines as an observational
tool, the likelihood of the existence of large quantities of molecular
gas in this situation, and the possibility that molecular gas could
constitute a large portion of the obscuring torus, theoretical models
of the molecular emission of the torus are fairly uncommon (see Wada
\& Norman 2002; Wada et al. 2009; P\'{e}rez-Beaupuits et al. 2011 for
examples of these simulations).

In this work, the models of Ballantyne (2008) are combined with \textsc{Ratran} (Hogerheijde \& van der
Tak 2000), a radiative transfer code, to estimate the CO line emission
from these objects.  Following this procedure, CO spectral line energy
distributions (SLEDs) are created, and brightness temperature ratios
calculated.  Section~\ref{sect:calc} contains the descriptions of the
nuclear starburst disk model, the molecular mass fraction model, and
\textsc{Ratran}.  Section~\ref{sect:Results} conveys our central
results and discusses their significance and physical origin, while
also comparing these results to observations of
starburst and AGN host galaxies.  In
Section~\ref{sect:DiscussionConclusion}, the findings of this work are
compared with the conclusions of similar simulations, and the
possibility for the detection of AGN-obscuring starburst disks is
explored.  Finally, in Section~\ref{sect:Conclusion}, concluding
remarks and analysis are presented.

A $\Lambda$-dominated cosmology is assumed in this paper, when
necessary.  The following parameters are used: $H_{\circ} = 71$ km
s$^{-1}$ Mpc$^{-1}$, $\Omega_{\Lambda}$ = 0.73, and $\Omega_{m}$ =
0.27 (Spergel et al. 2003; Spergel et al. 2007).

\section{Calculations}
\label{sect:calc}
\subsection{Review of the Starburst Disk Model}
\label{sect:AGNTorusModel}

The nuclear starburst disks are taken from the models of Ballantyne (2008), which are based on the one-dimensional, analytic model of Thompson et al. (2005). A short elucidation of the central tenets and assumptions of this model will be presented here.

The material of the disk is assumed to be a single phase medium. The various properties of the gas are calculated for discrete values of the distance, $r$, from the central SMBH.  The black hole mass is used to calculate the velocity dispersion of the stars in the galactic bulge, $\sigma$, according to the $M_{BH}$--$\sigma$ relationship (Ferrarese \& Merritt 2000; Gebhardt et al. 2000; Tremaine et al. 2002).   The gas is modeled as rotating at the Keplerian frequency at all radii in response to a gravitational potential composed of the sum of a point mass potential for the central black hole and the potential of an isothermal sphere to represent the galactic bulge.  Star formation is unresolved but calculated locally by assuming that the star formation rate behaves so as to maintain a Toomre's parameter of one, implying that the gas is locally marginally stable against self-gravity (i.e. $Q = \kappa_{\Omega} c_{s}/\pi G \Sigma_g = 1$, where $\kappa_{\Omega}$ is the epicyclic frequency, $\Omega$ is the Keplerian frequency, $\Sigma_g$ is the surface density of gas, and $c_s$ is the sound speed).  From this assumption, the density, $\rho$, may be explicitly calculated as a function of the Keplerian frequency, $\Omega$.  The vertical support of the disks results from the feedback of the star formation.  The radiation pressure of the infrared emissions given off by the dust forms the primary means by which this support is achieved.  Accretion is assumed to occur by means of a global mechanism for the shedding of angular momentum.  The process is assumed to allow the in-falling gas to achieve a radial velocity, $v_{r}$, equal to a constant fraction, $m$, of the speed of sound.  The input parameters of this model are the mass of the central black hole $M_{BH}$, the accretion efficiency parameter $m$, the gas fraction $f_{gas}$, the outer radius of the disk $R_{out}$, and the dust-to-gas ratio $d_{dtg}$.  The models use the dust opacities taken from the work of Semenov et al. (2003) with a density of $10^6$ cm$^{-3}$ and normal, homogeneous, and spherical grains.  Calculations were then undertaken for all permutations of the input parameters, as detailed in Ballantyne (2008).

In order to model the nuclear regions of AGN host galaxies, the nuclear starburst disks must be capable of feeding the AGN, in addition to providing the required obscuration (Thompson et al. 2005; Ballantyne 2008, Section 2.1).  Because of this requirement, the advection timescale, $\tau_{adv} = r/v_{r}$, must be less than the star formation timescale, $\tau_{*} = 1/(\eta \Omega)$, where $\eta$ is the star formation efficiency.   These considerations allow one to select the models most likely to fuel and obscure the central AGN by considering whether a region exists within the model that meets the following criteria: 1) a mid-plane temperature greater than 900 K and 2) a star formation rate greater than 10\% of the rate at $R_{out}$.  The first criterion ensures that the gas temperature reaches the dust sublimation temperature, creating a severe vertical opacity gradient.  The second allows for the exclusion of models that only marginally surpass the first criterion but are still largely incapable of the obscuration required by the unified model of AGN.  Those models that met both requirements were judged to contain a parsec-scale starburst capable of obscuring the central AGN.  Models satisfying neither requirement were found to lack an obscuring parsec-scale starburst region.  The models that failed to achieve a parsec-scale starburst had significantly lower temperatures than the starburst disk models, usually in the range of several tens of Kelvins and occasionally rising to as high as 200 K.  In total, approximately 41\% of the models tested met the starburst criteria (Ballantyne 2008, Section 2.2).  Starburst disks were produced by sets of parameters throughout the parameter space.  However, generally speaking, increasing values of all of the parameters, except the outer radius, were found to be most conducive to producing a potentially obscuring starburst.  The trend seen in $R_{out}$ indicates that smaller outer radii create conditions more amenable to obscuring starbursts (See Ballantyne 2008, Figure 3).  All models, regardless of classification, were found to be high density and high pressure environments, which, as mentioned in Section \ref{sect:intro}, are conditions favoring the formation of large quantities of molecular gas.

In order to utilize the \textsc{Ratran} program, a method of determining the molecular mass fraction in each annulus of the disk was needed. The densities at which molecular gas dominates over the atomic phase have been found to be relatively modest (i.e. approximately 100 cm$^{-3}$) compared to the densities present in the nuclear starburst disk models offered here (e.g., Solomon \& Wickramasinghe 1969; Solomon \& Vanden Bout 2005; Gnedin et al. 2009).  Also, it has been shown that the formation rate of molecular hydrogen drops to essentially zero for temperatures above approximately 1000 K, and collisional destruction eliminates any remaining H$_{2}$ in the area (Cazaux \& Tielens 2004; Pelupessy et al. 2006).  As mentioned above, the nuclear starburst disk model predicts extensive star formation in these high temperature regions, as well.  Because of the very high densities present in the disk models, ranging from approximately 500 cm$^{-3}$ at $R_{out}$ to $>10^{9}$ cm$^{-3}$ at the innermost radius of the disk, the molecular mass fraction, $f_{m}(r)$, was set to unity for regions of the disk with temperatures less than 900 K and to zero for the remaining portions of the disk\footnote{A more complicated molecular mass fraction recipe based on the static model of Pelupessy et al. (2006) was employed to test this scenario, and similar results were found.  The molecular mass fractions of the annuli of the disks were found to lie between 0.96 and 1.0 for all of the models analyzed.}.  The temperature, $T$, is taken directly from the mid-plane temperature of the nuclear starburst disk models.  After the molecular mass fraction is calculated, it is used to find the number density of molecular hydrogen from the density of the nuclear starburst disk model by assuming that all of the gas is hydrogen.  We then use a constant factor of $n_{CO}/n_{H_2} = 10^{-4}$ to determine the number density of carbon monoxide.  This value is frequently used as the upper limit on CO abundance (Klemperer 2006).

\subsection{Molecular Line Emission}
\label{sect:Ratran}
With the properties calculated in section \ref{sect:AGNTorusModel}, we employ the \textsc{Ratran} code written by Hogerheijde \& van der Tak (2000) to calculate the line emission of the disks.  We use the one-dimensional (spherically symmetric) form of the code.  For a review of the assumptions and techniques of this program, one may see Hogerheijde \& van der Tak (2000).

The inputs of the \textsc{Ratran} code are taken from the nuclear starburst disk models and the molecular mass fraction scheme detailed in Section~\ref{sect:AGNTorusModel}.  We assume that the gas at each radius, $r$, is well-mixed with the dust and, thus, that the molecular gas kinetic temperature is equal to the mid-plane temperature.  Furthermore, we set the dust temperature equal to this value because of the high density involved in the nuclear starburst disk models.  We approximate the turbulent line width by the local speed of sound, which is calculated by $c_s = \sqrt{P/ \rho}$, where $c_s$ is the speed of sound and $P$ and $\rho$ are the local pressure and density from the nuclear starburst disk model.  The Doppler broadening parameter, $b$, is calculated as $b = (2\sqrt{\ln(2)})^{-1} c_s$.  The radial velocity is also included in the inputs for the \textsc{Ratran} calculation and is calculated as follows: $v_{r} = m c_{s}$, as described in Section~\ref{sect:AGNTorusModel} and Ballantyne (2008).

As the nuclear starburst disk models depend explicitly on the absorption and emission of dust to explain the vertical support of obscuring disks, a dust opacity is included in the \textsc{Ratran} calculation.  As mentioned above, Ballantyne (2008) used the Semenov et al. (2003) Rosseland mean opacity curve.  However, for the \textsc{Ratran} code, we require a frequency-dependent opacity.  Therefore, we include a broken power law dust opacity based on the work of Pollack et al. (1994) with the following form: $\kappa \propto \lambda^{-\beta}$, where $\beta$ is the dust opacity spectral index.  Semenov et al. (2003) reference this work, utilize the same distribution of dust components, and employ similar techniques to calculate the properties desired.  The wavelengths of the emission of the lowest seventeen rotational lines of CO range from 2.6 mm to approximately 153 $\mu m$.  Table 4 of Pollack et al. (1994) includes spectral indices for wavelengths between 650 $\mu m$ and 2.3 mm for a wide variety of models of dust composition.  We employ these indices for the wavelength range over which they apply.  We then measure indices for the range between 100 $\mu m$ and 650 $\mu m$ from Figures 2b and 2c of Pollack et al. (1994).  To attain self-consistency with the Ballantyne (2008) calculations, we use spectral indices for spherical particles composed of segregated materials.  Segregation in the Pollack et al. (1994) model corresponds to the condition of homogeneity in the Semenov et al. (2003) model, namely, that the dust particles are composed of a single constituent, rather than being a conglomeration of all the substances constituting dust.  Also, for the sake of consistency, we set the dust opacity equal to zero for the regions in which the mid-plane temperature exceeds the dust sublimation temperature.  We select a particle radius of 3 $\mu m$, well within the size distribution of Semenov et al. (2003).  Armed with these meditations, we select a dust model with a temperature of 700 K for the nuclear starburst disk models that achieved a parsec scale starburst or are borderline cases.  A dust model of temperature 100 K is used for models that did not meet either criterion.  Lastly, the opacities in Table 4 of Pollack et al. (1994) are given in terms of total disk mass.  However, the \textsc{Ratran} program requires that they be in terms of total dust mass.  Thus, the following simple conversion is made: $\kappa_{M_{dust}} = \kappa_{M_{tot}} \times (M_{tot})/(M_d) = \kappa_{M_{tot}} (\delta_{gtd} + 1)$, where $M_{dust}$ is the total mass of dust, $M_{tot}$ is the total mass of the disk, and $\delta_{gtd}$ is the absolute value of the gas-to-dust ratio.  Assuming a local ISM value for the gas-to-dust ratio of 150 (Draine \& Lee 1984; Young \& Scoville 1991), the absolute value of the gas-to-dust ratio, $\delta_{gtd}$, is related to the dust-to-gas ratio parameter of the Ballantyne (2008) disk models, $d_{dtg}$, as follows : $\delta_{gtd} = 150/d_{dtg}$.  The spectral indices and opacities used in our calculations are summarized in Table \ref{tab:dust}.
The number of channels and the channel width used for the \textsc{Ratran} calculations are 8000 and 0.05 km s$^{-1}$, respectively, well within the capabilities of ALMA (Schieven 2011; Vila Vilaro 2011).

\begin{deluxetable}{lcccc}
\tabletypesize{\small}
\tablewidth{0pt}
\rotate
\tablecaption{The spectral indices and normalizations for the power law dust opacity models\label{tab:dust}}
\tablehead{\colhead{Model Type} & \colhead{$\beta(100-650 \mu m)$} & \colhead{$\kappa(650~\mu m)$ (cm$^2$ g$^{-1}$)} & \colhead{$\beta(650-2700 \mu m)$} & \colhead{$\kappa(1~\mathrm{mm})$ (cm$^2$ g$^{-1}$)}}
\startdata
Starburst & 1.83 & $3.7 \times 10^{-2}$ & 0.98 & $2.4 \times 10^{-3}$ \\
Failed & 2.73 & $7.1 \times 10^{-2}$ & 1.46 & $3.8 \times 10^{-3}$ \\
\enddata
\tablecomments{Given here are the values for the spectral indices and normalizations used in the dust model detailed in Section~\ref{sect:Ratran}.  A broken power law, $\kappa \propto \lambda^{-\beta}$, is employed with one index serving for wavelengths between 100 and 650 $\mu m$ and a second for wavelengths between 650 $\mu m$ and 2.7 mm.  The opacity power law normalizations are given for 650 $\mu m$ and 1 mm.  Note that the opacity normalization constants are given in terms of total disk mass, as detailed in the text.}
\end{deluxetable}

Following the completion of the \textsc{Ratran} program, the brightness temperature maps produced are analyzed using the Miriad software package (Sault et al. 1995).  In particular, dust emission is removed from the line images by subtracting the first channel from all the subsequent channels for each image.  The images are then convolved to a beam with a full-width at half maximum (FWHM) , $\theta$, appropriate for ALMA given by the following equation: $\theta = (76)/(D \nu_{obs})$ arcsecs, where $D$ is the maximum baseline of the configuration used in kilometers and $\nu_{obs}$ is the frequency at which observations are being made in GHz (Schieven 2011).  For our calculations, we use a value of $D = 0.125$ km, the maximum baseline for the most compact configuration available.  Following convolution to the desired spatial resolution, the velocity-integrated intensity, $I_{CO}$, is found using the Miriad \emph{moment} function.  The velocity-integrated intensity is then multiplied by the source surface area to get the luminosity in K km s$^{-1}$ pc$^{2}$, $L_{CO}^{\prime}$:
\begin{equation}
\label{eq:GeoLum}
L_{CO}^{\prime} = 4\pi \left((R_{out})^2 - (R_{in})^2\right) I_{CO},
\end{equation}
where $R_{in}$ and $R_{out}$ are the inner and outer radii of the disk, respectively, and have units of parsecs and $I_{CO}$ has units of K km s$^{-1}$. $R_{in}$ is, generally, two to three orders of magnitude smaller than $R_{out}$ and, thus, was neglected in all computations.  This luminosity was then converted into conventional units and used to find a flux as follows:
\begin{equation}
\label{eq:PapaLum}
S_{line} = \frac{2 k_{B} \nu_{\circ}^{3}}{D_{L}^{2} c^3} L_{CO}^{\prime},
\end{equation}
where $D_{L}$ is the luminosity distance, $k_{B}$ is the Boltzmann constant, $c$ is the speed of light, and $\nu_{\circ}$ is the rest frequency (Papadopoulos et al. 2010a; equation 5).

\section{Results}
\label{sect:Results}
Using equations \ref{eq:GeoLum} and \ref{eq:PapaLum}, brightness
temperature ratios and CO spectral line energy distributions (SLEDs)
are calculated for eighteen nuclear starburst disk models (see Table 2).
The particular models chosen are selected to sample the parameter
space of the nuclear starburst disk model widely enough that the
variation of the CO emission with the input parameters can be
discerned.  The set of models analyzed here includes an equal number of
failed and starburst cases.  


\subsection{The CO SLED from an Ultra-Compact Starburst}
\label{sub:explanation}
The solid black line in Figure 1 shows an example flux SLED produced by
an ultra-compact starburst with an outer radius of 250~pc at $z=0.8$, a typical redshift for the
obscured Seyfert galaxies that dominate the X-ray background (e.g.,
Ueda et al.\ 2003). The input parameters of this starburst model are listed
in Table 2, but its flux SLED shape and magnitude are typical for all
the studied starburst models. The flux SLED shows two nearly symmetric
features: a
strong emission bump peaking at $J_{\mathrm{Upper}} \approx 6$--$7$,
and a deep absorption trough that reaches a minimum at
$J_{\mathrm{Upper}} \approx 14$--$15$. To understand the origin and
physics behind these features we consider the flux SLEDs of the disk
truncated at various radii. The red, dotted line with open triangles
shows the SLED produced from only the inner 200~pc of the disk and is
very similar to the total SLED, albeit with somewhat less
absorption. In contrast, the SLED from the inner 50~pc of the disk
(the red, dotted line with open pentagons) shows a significantly
different shape with both the emission and absorption peaks shifted to
higher values of $J_{\mathrm{Upper}}$, indicating significant amounts
of highly excited CO consistent with the high temperature and
pressures predicted in this region of this disk (e.g., Ballantyne
2008). The density and temperature continues to rise culminating in a
significant burst of star-formation in the inner 10-pc of the
disk. The CO flux SLEDs from the inner 1.4, 2.5 and 6.3~pc of the starburst
disk are also shown in Figure~1, but they all overlap and have line
fluxes close to zero at all values of $J_{\mathrm{Upper}}$, indicating
that these regions produce a featureless blackbody continuum, as
expected for an optically thick mixture of dust and gas. 

Figure 1 shows that the innermost portions of the
 ultra-compact starburst disk do not contribute substantially either to the
 emission or the absorption features seen in the final CO SLED from the entire
 disk. Recall that the densities in these models are universally high
 ranging from $\sim 500$~cm$^{-3}$ at the outer radius to $\sim
 10^{9}$~cm$^{-3}$ in the innermost portions of the disk (e.g.,
 Thompson et al.\ 2005; Ballantyne 2008). The fact that the critical
 densities of CO lines tend to be fairly low with respect to most of the densities in the disk means that CO should be
 fairly well excited throughout the disk. Therefore, we find that,
 overall, the disk acts similarly to a star: there is a hot and dense
 central region that produces a blackbody continuum which then must
 propagate through a large column of material with a steep density
 and temperature gradient. Intervening molecular gas, being always cooler
 than the gas interior to it, will absorb the higher-$J_{\mathrm{Upper}}$ lines while still emitting substantially at the
 lower-$J_{\mathrm{Upper}}$ line because of the relatively low critical
 densities of CO lines. The final CO SLED will therefore have both
 emission and absorption features due to the unique strong density and
 temperature profiles predicted by the ultra-compact starburst disk
 model. The emission lines highlight the excited nature of the outer
 molecular layers, while the absorption features indicate the presence of
the hot, high-pressure center that is buried deep inside the disk.

The above physical reasoning predicts that there will not be
strong differences in the CO SLEDs predicted by starburst models and
those that failed to produce a pc-scale burst because the density profiles of both classes are similar. However, although the innermost portions of
 the disk may not contribute substantially to the total flux in any
 given line, they must affect the energy balance of the disk, as the
 failed models (by definition) do not achieve the large temperatures found in the inner regions of the starburst models. Therefore, we
 expect, and the data confirm (see below), that the CO SLEDs from
 failed models are shifted to lower $J_{\mathrm{Upper}}$ and are less
 luminous than ones produced by ultra-compact starbursts.

Finally, we emphasize that these calculations assume spherical
symmetry, so the lines-of-sight always pass through the `edge' of the
starburst disk, and will therefore observe the maximum of the
radiative transfer effects. Other, more realistic, lines-of-sight may
predict weaker emission and absorption features in the SLED. This will
be the subject of future work. 

\begin{figure}
\label{fig:opacity}
\includegraphics[scale=0.4,angle=270]{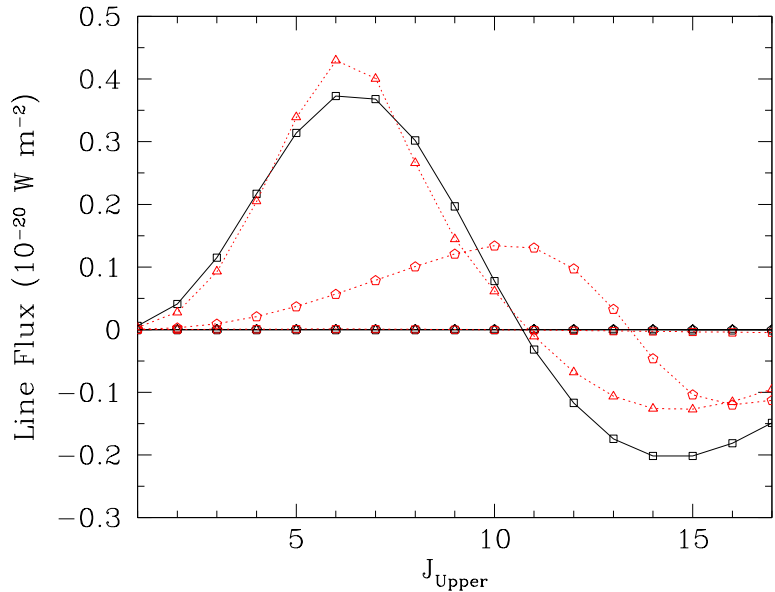}
\caption{CO flux SLEDs (at $z=0.8$) showing the contributions to the
  SLED from different portions of a nuclear starburst disk (see marked starburst model in Table
  2).  The black, solid line with square markers is the flux SLED of
  the entire model with an outer radius of 250 parsecs. The red,
  dotted line with triangles is the SLED of the inner 200 pc of the
  disk model. The red, dotted line with pentagons is the SLED for the
  region with an outer radius of 50 pc.  The contributions to the SLED
  from the inner 1.4, 2.5 and 6.3 pc all lie along the zero flux
  line. Thus, much like a star, the outermost regions of the disk
  determine the overall shape of the flux SLED, while the innermost
  and hottest portions of the disk affect the magnitude of the SLED.}
\end{figure}

\subsection{Distribution of CO SLED Properties}
\label{sub:histo}

\begin{figure}
\label{fig:rathis}
\includegraphics[scale=0.4,angle=270]{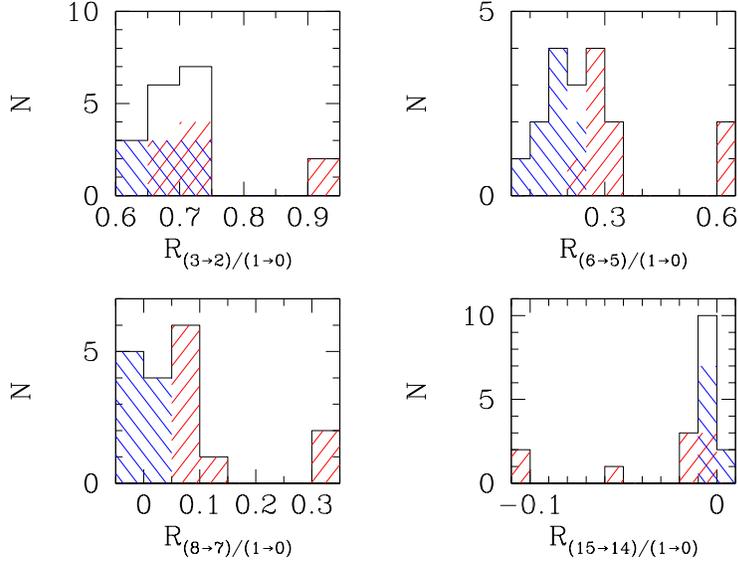}
\caption{Histograms of four brightness temperature ratios for the
  eighteen models analyzed in this work.  The distribution of ratios
  for the starburst models with pc-scale bursts is shaded with
  right-leaning (red) diagonal lines.  The distribution for the failed
  models is marked by left-leaning (blue) lines.  Note that the
  starburst models generally have higher values for the brightness
  temperature ratios for lines in which most models have not fallen
  into absorption.  For lines in which the majority of models show
  absorption, starburst-classified models tend to have stronger
  absorption.  Note also that the separation between the distributions
  of failed and starburst models tends to increase from the histogram
  of R$_{(3\rightarrow 2)/(1\rightarrow 0)}$ in which a great overlap
  exists to the distribution for R$_{(8\rightarrow 7)/(1\rightarrow
    0)}$ in which no overlap between the two classes occurs.  For
  ratios involving the very highest lines explored, such as $J =
  15\rightarrow 14$, little separation occurs because most models show
  little emission or absorption at these high lines. If the nuclear
  starforming region can be resolved, these ratio differences between
  the two model classes could be used to determine if ultra-compact,
  obscuring starbursts exist in the observed galaxies.}
\end{figure}

Turning now to the SLED properties of the eighteen analyzed models,
Figure 2 displays histograms of four brightness temperature ratios,
R$_{(3\rightarrow 2)/(1\rightarrow 0)}$, R$_{(6\rightarrow
  5)/(1\rightarrow 0)}$, R$_{(8\rightarrow 7)/(1\rightarrow 0)}$, and
R$_{(15\rightarrow 14)/(1\rightarrow 0)}$.  Distributions for those
models classified as starbursts are shaded with right-leaning (red)
lines, while those of the failed models are shaded with left-leaning
(blue) lines.  The average ratios for all of the models analyzed here,
as well as for the starburst and failed classes separately, can be
found in Table 3.  First, as expected from Sect.~3.1, Table 3 shows
that the average brightness temperature ratios for the starburst
models are considerably higher in magnitude than those of the failed
models, indicating that the models containing an obscuring
parsec-scale starburst possess a much higher excitation of CO lines
than those that fail to meet this requirement. There is also an
increasing separation between the distributions of the two classes as
one analyzes the R$_{(3\rightarrow 2)/(1\rightarrow 0)}$,
R$_{(6\rightarrow 5)/(1\rightarrow 0)}$, and R$_{(8\rightarrow
  7)/(1\rightarrow 0)}$ histograms in sequence.  This distinguishing
feature allows for the possibility that temperature ratios could be
used to search for obscuring parsec-scale starburst regions, if these
regions can be resolved within their host galaxies.  Finally, note
that absorption is seen in most models for transitions with
$J_{\mathrm{Upper}}$ greater than or equal to approximately 10.  As can be seen
in the histogram for R$_{(15\rightarrow 14)/(1\rightarrow 0)}$, and
expected from Sect.~3.1, starburst models tend to have stronger absorption (i.e. more negative temperature ratios).  This feature is even more apparent for lines with $J_{\mathrm{Upper}}$ somewhat lower than 15, such as the ratio of the $J = 13\rightarrow 12$ line to the $J = 1\rightarrow 0$ line. 

\subsection{Comparison to Observed CO SLEDs}
\label{sub:compare}
Figure 3 shows luminosity SLEDs for two of the eighteen models
analyzed in this work.  The starburst model is marked by a black,
solid line, while the failed model is depicted by a black, dotted
line.  The input parameters of both models are identified in Table 2.
The observed SLEDs of various galaxies are offered for comparison,
including the inner 800~pc of the nearby starburst galaxy M82 (Sanders
et al. 2003; Ward et al. 2003; Panuzzo et al. 2010; magenta, solid
line with open squares), the nearby ULIRG Mrk 231 (van der Werf et
al. 2010; red, long-dashed line with open pentagons), the nearest
ULIRG Arp 220 (Rangwala et al. 2011; green, short-dashed line with
open triangles), and the z=2.958 ULIRG HERMES J105751.1+573027
(Conley, et al. 2011; Scott et al. 2011; blue, dotted line with filled
circles).  Comparing the SLEDs of the two nuclear starburst disk
models, one notes that the magnitudes of the luminosities are higher
for the successful starburst model, while the failed model is
generally less luminous.  Also, the ratio of the luminosities of the
higher lines to the lowest transition tend to be lower for the failed
models than for the starburst models, indicating a more substantive
excitation in the successful parsec-scale starburst models (see also
Sect. 3.1, Figure 2 and Table 3).  Lastly, note that the failed model falls into
absorption (i.e. its luminosity SLED ends) at a lower $J_{\mathrm{Upper}}$
value than the starburst model.

The line luminosities estimated in this work are comparable to those
reported in the literature if the size of the emission region is taken
into account.  The disk model of Thompson et al. (2005) was developed
to describe LIRGs, ULIRGs, and other extreme starburst environments,
and, thus, we compare our emission luminosities to observations for
these objects, in addition to quasars and AGNs.  Several reports in
the literature on ULIRGs and quasar host galaxies find CO line
luminosities for the $J = 1 \rightarrow 0$ transition in the range of
10$^{9-11}$ K km s$^{-1}$ pc$^{2}$ (e.g. Chapman et al. 2008; Braun et
al. 2011; Riechers 2011; Riechers et al. 2011; Scott et al. 2011; Wang
et al. 2011).  In particular, the work of Smol\v{c}i\'{c} \& Riechers
(2011) describes observations of CO line emission of various nearby (z
$\leq$ 0.1) AGN.  The values reported in Table 3 of Smol\v{c}i\'{c} \&
Riechers (2011) track the range of CO line luminosities found in this work for the lowest lines.  Therefore, the magnitude of the luminosities
found in this work seems reasonable.  One may also compare the
luminosities of the models produced here to those of the observed
galaxies given in Figures 3.  This comparison sustains our
previous conclusion that the models are an order of magnitude or so
less luminous because of their substantially smaller masses of emitting material.  The fact that the emission of the host galaxy of the
nuclear starburst disk could be comparable to or greater than
the emission of the disk, itself, implies that the emission of the
innermost regions of the galaxy may have to be isolated in order to observe
them.  However, the observations plotted
in Figure 3 are from galaxies selected to have substantial star formation
rates and be very luminous in CO and will not necessarily be
representantive of the galaxies that host obscured Seyferts at $z < 1$,
which typically have much lower rates of star formation (e.g.,
Silverman et al.\ 2009; Pierce et al. 2011). It would be interesting
to compare the predicted SLEDs against ones obtained from
X-ray selected $z<1$ AGNs.

For better relative comparisons of the shape of the model CO SLEDs to
observations, Figure 4 presents normalized flux (top panel) and
luminosity (bottom panel) SLEDs for the models and the galaxies shown
in Figure 3.  The CO luminosity SLEDs are converted to solar
luminosities (Papadopolous et al. 2010a; eq. 5) and are
normalized to the total infrared luminosity between 8 and 1000 $\mu m$.  The calculation of
the infrared luminosity for the starburst disk models is detailed in Ballantyne
(2008).  The starburst disk models and the observed galaxies seem to have similar
luminosity SLEDs, when normalized to the total infrared luminosity,
indicating that, for a given reservoir of dusty star-forming gas, the
ultra-compact starbursts produce a similar amount of CO luminsity for
$J_{\mathrm{Upper}} \la 9$. However, at higher values of $J_{\mathrm{Upper}}$,
  the models predict absorption, which is not observed in the
  large-scale emission of star-forming galaxies.

The observed flux SLEDs (Fig 4; top) have a shape generally peaking
in the range of $J$ = 6--8 similar to the starburst model peak at $J$= 6--7, while the failed model reaches a maximum at approximately
5. In addition, the observed flux SLEDs tend to be flatter at high
$J_{\mathrm{Upper}}$ than the SLEDs of the models, possibly indicating the
presence of an X-ray dominated region (XDR; see, e.g., van der Werf et
al. 2010). Feedback from X-ray heating is not included in our
calculations, but we show in Section~\ref{sect:XDR} that it would
impact the very inner-most region of a starburst disk, which has the
least impact on the shape of the CO SLED (Figure 1). Crucially, none of the observed flux SLEDs show
any indication of falling into absorption at large
$J_{\mathrm{Upper}}$. As the predicted absorption feature is a natural
outcome of the steep density and temperature gradients within the
starburst (see Sect. 3.1) and would not be altered by a presence of a XDR, we conclude that
the absorption portion of the SLED is the most obvious and important
observational prediction of the model. However, as discussed below,
careful observational strategies and techniques (e.g., stacking) will
be required to search for this signature.

\begin{figure}
\label{fig:lum}
\includegraphics[scale=0.4,angle=270]{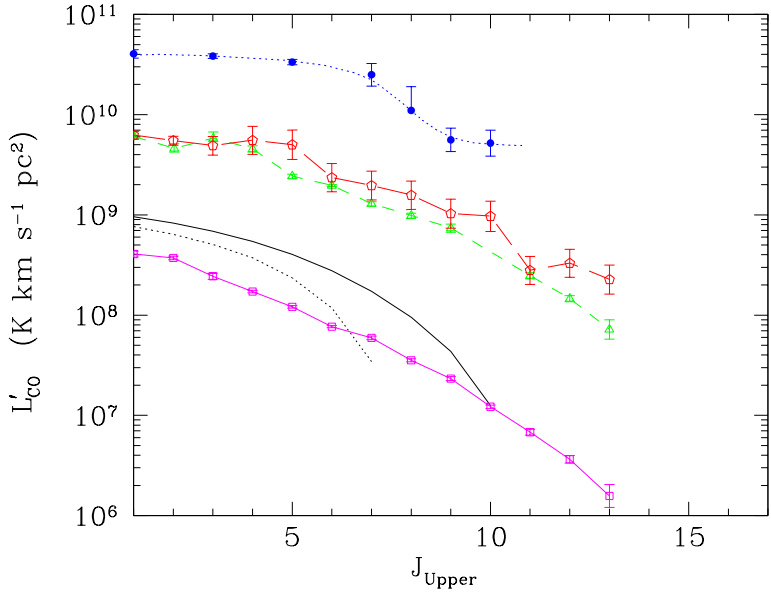}
\caption{Plot of CO luminosity SLEDs for a starburst and a failed
  model, both of which are identified in Table 2.  The starburst model
  is marked by a black, solid line, while the model without a
  parsec-scale obscuring starburst is marked by a black, dotted
  line. The CO SLED of the inner 800~pc of M82, a nearby starburst galaxy, is marked by solid, magenta line with open, square markers (Sanders et al. 2003; Ward et al. 2003; Panuzzo et al. 2010).  Mrk 231, a nearby ULIRG, is depicted by a red, long-dashed line with open, pentagonal markers (van der Werf, et al. 2010).  The nearest ULIRG Arp 220 is plotted with a green, short-dashed line and open, triangular markers (Rangwala, et al. 2011).  Lastly, a distant (z = 2.958) ULIRG HERMES J105751.1+573027 is marked by a blue, dotted line with filled, circular markers (Conley et al. 2011; Scott et al. 2011).}
\end{figure}

\begin{figure}
\label{fig:flux}
\includegraphics[scale=0.46,angle=270]{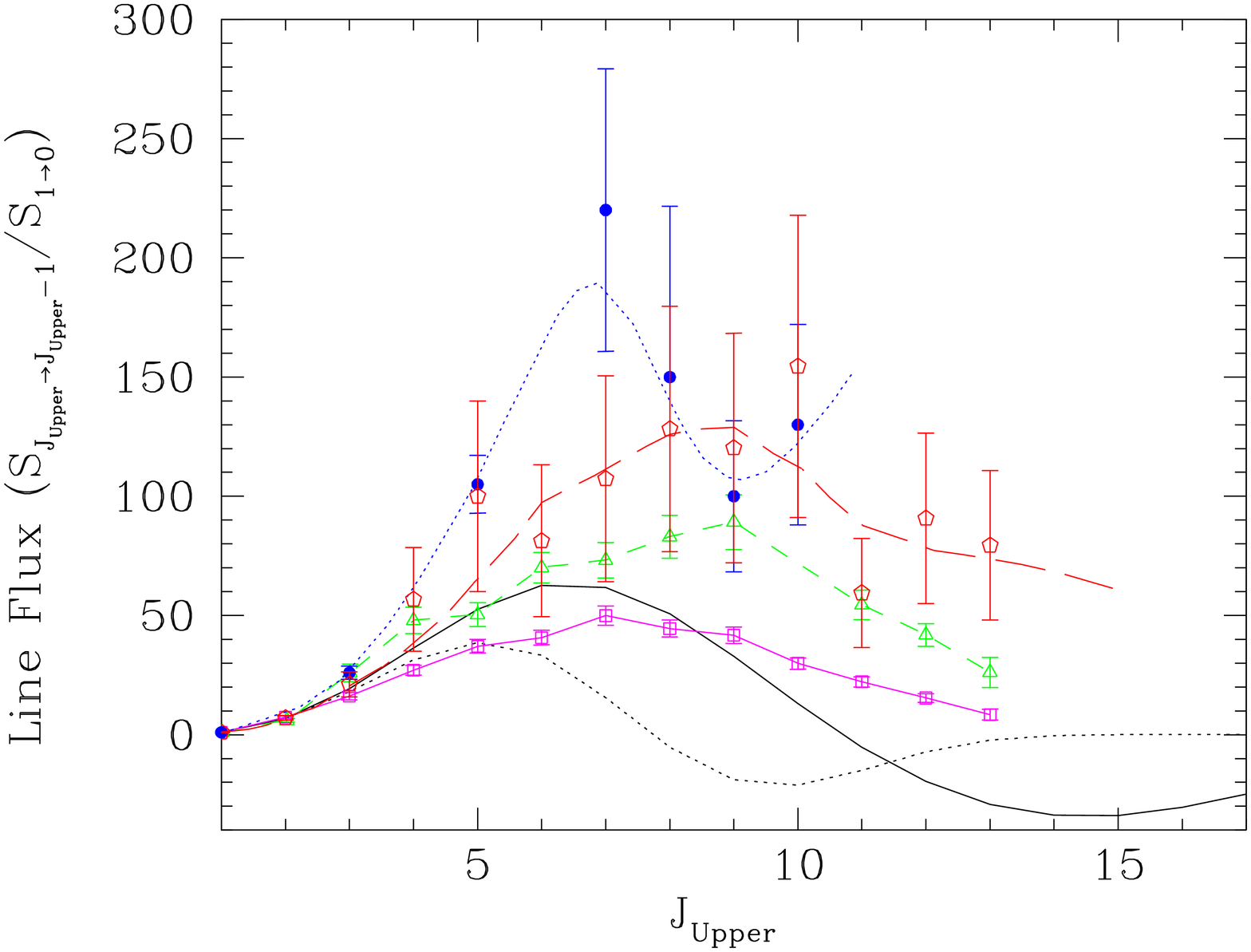}
\includegraphics[scale=0.46,angle=270]{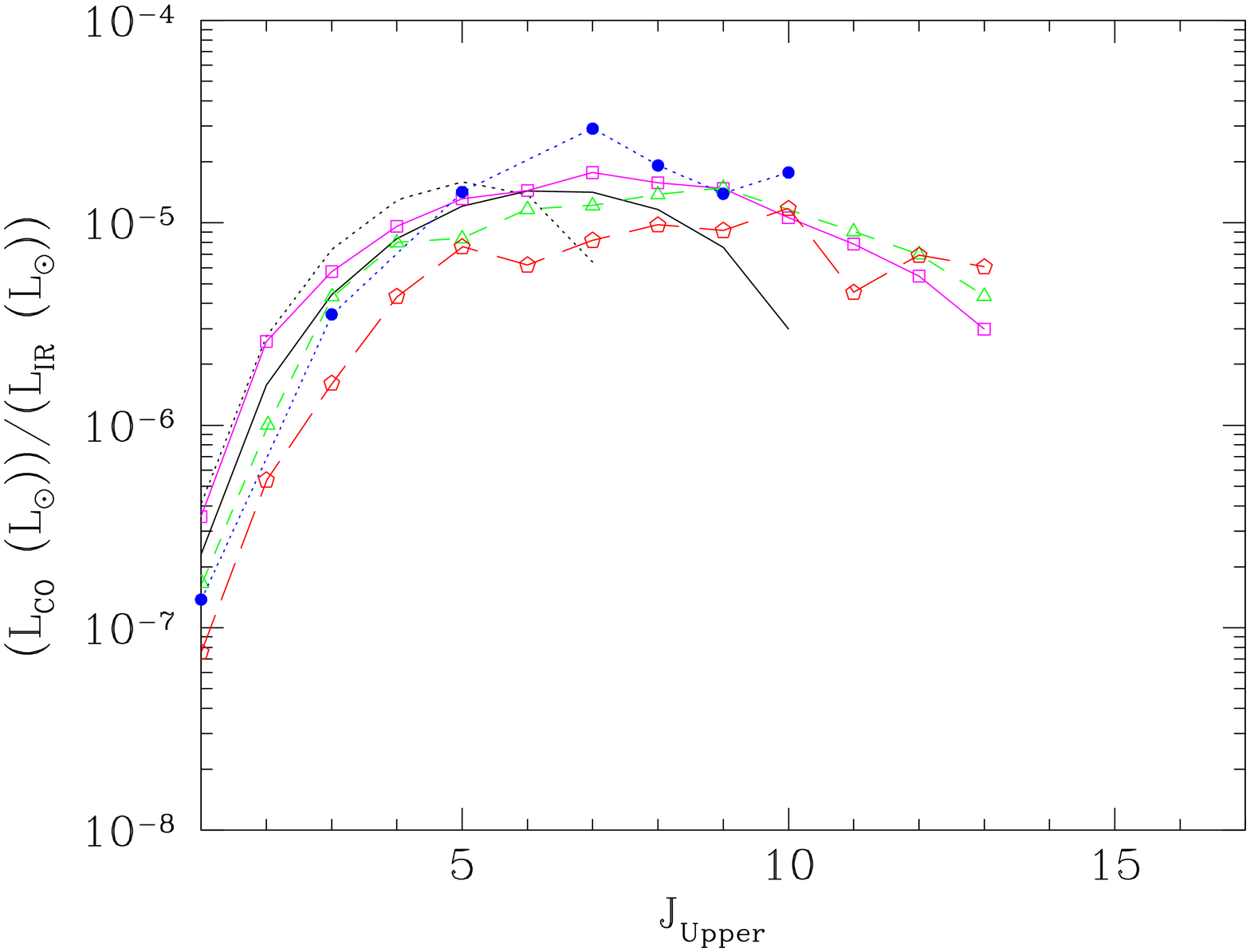}
\caption{(Top) Plot of the CO flux SLEDs for a starburst and a failed
  model, both of which are identified in Table 2, as well as the
  observed SLEDs for several galaxies.  The distributions are
  normalized by the flux of the lowest ($J = 1\rightarrow 0$)
  transition. (Bottom) Plot of CO luminosity SLEDs, normalized by the
  total infrared luminosity (8-1000 $\mu m$), for the models and
  observed galaxies depicted in the other panel, as well as Figure
  3. All luminosities are in solar units. The identification and marking scheme of the plot is identical to that of Figure 3.}
\end{figure}

\begin{deluxetable}{cccccl}
\tabletypesize{\small}
\tablewidth{0pt}
\tablecaption{Model Input Parameters\label{tab:mod}}
\tablehead{ \colhead{$\frac{M_{BH}}{M_{\odot}}$} & \colhead{$d_{dtg}$} & \colhead{$m$} & \colhead{$R_{out}$ (pc)} & \colhead{$f_{gas}$} & \colhead{Model Class}}
\startdata
7.5 & 5.0 & 0.0750 & 250 & 0.10 & f \\
7.5 & 5.0 & 0.0750 & 250 & 0.90 & sb \\
7.5 & 10.0 & 0.0750 & 250 & 0.50 & sb \\
7.5 & 5.0 & 0.0100 & 250 & 0.50 & f \\
7.5 & 5.0 & 0.0250 & 250 & 0.50 & f \\
7.5 & 5.0 & 0.1000 & 250 & 0.50 & sb \\
7.5 & 1.0 & 0.0750 & 150 & 0.50 & f \\
7.5 & 1.0 & 0.0750 & 200 & 0.50 & f \\
7.0 & 5.0 & 0.0750 & 250 & 0.50 & f \\
8.0 & 5.0 & 0.0750 & 250 & 0.50 & sb \\
8.5 & 5.0 & 0.0750 & 250 & 0.50 & sb \\
7.5 & 1.0 & 0.0750 & 250 & 0.50 & f \\
7.5 & 5.0 & 0.0075 & 250 & 0.50 & f \\
7.5 & 1.0 & 0.0750 & 50 & 0.50 & sb \\
7.5 & 5.0 & 0.0500 & 250 & 0.90 & sb \\
8.0 & 5.0 & 0.0075 & 50 & 0.90 & sb \\
7.5 & 5.0 & 0.2000 & 250 & 0.50 & sb* \\
7.5 & 5.0 & 0.0500 & 250 & 0.50 & f* \\
\enddata
\tablecomments{Presented here are the input parameters and
  classifications for the eighteen models analyzed in this work. The
  models shown in Figures~1, 3 and 4 are indicated with asterisks.  Lastly, the model class f stands for failed, which indicates that the model in question does not contain a parsec-scale obscuring starburst, and sb represents starburst, which is used to label all models that meet the starburst criteria of Section~\ref{sect:AGNTorusModel}.}
\end{deluxetable}

\begin{deluxetable}{lccccc}
\tabletypesize{\small}
\tablewidth{0pt}
\tablecaption{Brightness Temperature Ratios\label{tab:ratios}}
\tablehead{\colhead{Model Type} & \colhead{R($\frac{3\rightarrow 2}{1\rightarrow 0}$)} & \colhead{R($\frac{6\rightarrow 5}{1\rightarrow 0}$)} & \colhead{R($\frac{6\rightarrow 5}{3\rightarrow 2}$)} & \colhead{R($\frac{8\rightarrow 7}{1\rightarrow 0}$)} & \colhead{R($\frac{15\rightarrow 14}{1\rightarrow 0}$)}}
\startdata
failed & 0.66 & 0.15 & 0.23 & $-1.0\times 10^{-2}$ & $-2.6\times 10^{-5}$ \\
starburst & 0.71 & 0.29 & 0.41 & 0.10 & $-1.0\times 10^{-2}$ \\
\cutinhead{Averages}
$\langle$total$\rangle$ & 0.73 & 0.27 & 0.35 & $7.6\times 10^{-2}$ & $-2.9\times 10^{-2}$ \\
$\langle$starburst$\rangle$ & 0.78 & 0.37 & 0.45 & 0.15 & $-5.7\times 10^{-2}$ \\
$\langle$failed$\rangle$ & 0.68 & 0.17 & 0.24 & $9.7\times 10^{-4}$ & $-5.4\times 10^{-4}$ \\
\enddata
\tablecomments{Shown here are brightness temperature ratios for the models in Figures 3 and 4, as well as average ratios for all eighteen tested models, all of the successful starburst models, and the models that failed to achieve a starburst.  Note that the starburst ratios are significantly higher than those of the failed models, particularly when considering higher-J lines.}
\end{deluxetable}

\section{Discussion}
\label{sect:DiscussionConclusion}
\subsection{The CO Emission from Nuclear Starburst Disks}
\label{sect:properties}
This paper has explored the CO emission and absorption properties of the Ballantyne (2008) nuclear
starburst disk model. Eighteen models, nine of which
produced the crucial pc-scale starburst that can act as an obscuring
torus, were examined in detail, and found to produce CO luminosities and SLEDs similar to other
extragalactic sources (e.g., ULIRGs and starburst galaxies; see
Figures~3 and 4). Moreover, the shape of the CO SLEDs of the nuclear starburst disks are sensitive to the
presence or absence of the pc-scale starburst (Figures 2, 3, and 4 and Table 3)
with many of the predicted brightness temperature ratio differences exceeding
15\% depending on the existence of the pc-scale burst. Therefore, if the
nuclear starburst CO emission can be distinguished from that produced by
the host galaxy, the SLED will provide a direct test of the nuclear
starburst model for obscuring $z < 1$ Seyferts and provide help in understanding the coincidence of the peaks of the distribution of obscured AGNs and
the star formation history around $z \sim 1$  (e.g. Tozzi et al. 2001;
Barger et al. 2002; Hopkins 2004; Barger et al. 2005; Ballantyne
2008).

A potentially vital observational signature of the ultra-compact
starburst model SLEDs shown in Figures~1 and 4 is the occurrence of
absorption in the highest transitions. If confirmed, this absorption
would be a characteristic feature of nuclear starburst disks (see
Sect. 3.1); however,
its possible that this feature may be altered by X-ray heating by the
AGN. For example, Figure 4 shows an enhancement in the highest CO lines of Mrk 231, which are believed to result
from the influence of an AGN-powered XDR (van
der Werf et al. 2011). The possible effects of X-ray heating on the
calculated SLEDs are discussed further below.

\subsection{Comparisons with Previous Work}
\label{sect:Comparisons}
Wada \& Norman (2002) investigated three-dimensional, time-dependent
hydrodynamical simulations of a star-forming disk of material around a central
supermassive black hole where turbulent velocities
generated by supernova explosions provide the vertical support for the
flared torus of obscuring material (see also Wada et al.\
2009). P\'{e}rez-Beaupuits et al.\ (2011) 
present predictions for the relationship between CO luminosity and molecular
mass in this scenario, but do not present integrated flux
and/or luminosity SLEDs for the disk that could be compared with our
results.  However, it is clear that the two approaches are
complimentary with the high resolution modeling of the
P\'{e}rez-Beaupuits et al.\ (2011) simulations valuable for comparison to observations of nearby
galaxies, while the work presented here with its results, which are
integrated over frequency and space, is of more value for poorly
resolved and more distant AGN host galaxies.

\subsection{Detecting Nuclear Starbursts Around AGN}
\label{sect:Detection}
As shown in Sect.~\ref{sect:Results}, the nuclear starburst disks
predict CO luminosities large enough that these objects should be
detectable by current instruments. However, if the host galaxy is also
undergoing significant star-formation, the larger gas mass in the
galaxy will produce a CO SLED that could swamp the one from the
nucleus (Figures~3 and 4). One way to mitigate this problem is to
directly resolve the nuclear starburst disk, requiring an angular
resolution sufficient to resolve linear distances of $\sim 100$~pc at
$z \la 1$. The needed resolution may be possible for certain objects with
the fully operational ALMA in the most extended configuration with a maximum baseline of 16
km.  If one places the most luminous of the 18 starburst disks treated here (one
with a successful pc-scale starburst) at $z = 0.3$, the
$J = 3\rightarrow 2$, $J = 4\rightarrow 3$, and $J = 5\rightarrow 4$
lines exceed the $1\sigma$ sensitivity limit of ALMA and will be imaged
with angular resolutions that correspond to linear distances of a few
tens of parsecs. If this source is at $z = 0.2$, the $J = 2\rightarrow 1$ and $J =
3\rightarrow 2$ lines are greater than the $3\sigma$ limit and may
be resolved with an angular resolution of a few tens of parsecs. The $J = 5\rightarrow 4$ line rises above
the $5\sigma$ sensitivity limit at $z=0.2$, while maintaining a linear
resolution of approximately 32 parsecs, and the $J = 1\rightarrow 0$ and
$J = 7\rightarrow 6$ lines exceed the $1\sigma$ sensitivity
limit. Finally, at $z = 0.1$, the transitions from $J = 1\rightarrow 0$ to $J =
4\rightarrow 3$ are all greater than the $6\sigma$ detection limit
with angular resolutions in the tens of parsecs range, and the $J =
6\rightarrow 5$ and $J = 8\rightarrow 7$ lines exceed the $8\sigma$ limit.

For AGNs at $z \ga 0.3$, many of the challenges in directly detecting a nuclear
starburst disk could be overcome by leveraging information from other wavelengths, such as
selecting a sample of AGNs with high-resolution radio imaging or
mid-infrared colors that
have low to weak galactic-scale star formation. The nuclear starburst
disks would be expected to dominate the CO SLEDs of these galaxies.
To illustrate this, Figure 5 plots the CO SLED of the
Milky Way as measured by Fixsen et al.\ (1999) (green solid line),
plus the previously discussed starburst disk SLEDs from
Figures~3 and 4. Adding the Milky Way and the ultra-compact starburst SLED yields the solid black line, while the sum of the Milky Way and
the failed model gives the SLED shown by the dotted black line. Thus,
for galaxies with galactic-scale star-formation rates similar to those of the
Milky Way, the presence of a nuclear starburst disk would dominate the
observed CO SLED for $J_{\mathrm{Upper}} \la 8$. As discussed in
Sect.~3.1, the predicted absorption at higher values of
$J_{\mathrm{Upper}}$ is a direct consequence of the structure of the
compact starburst disk, and we expect that this feature will be
similarly dominate when observed against the background of a weakly
starforming galaxy. Figure 5 also indicates that stacking of CO SLEDs
from a sample of $z \la 1$ X-ray selected AGNs will be a useful
strategy to pull out the tell-tale signatures of the nuclear
starburst disk.

\begin{figure}
\label{fig:stack}
\includegraphics[scale=0.4,angle=270]{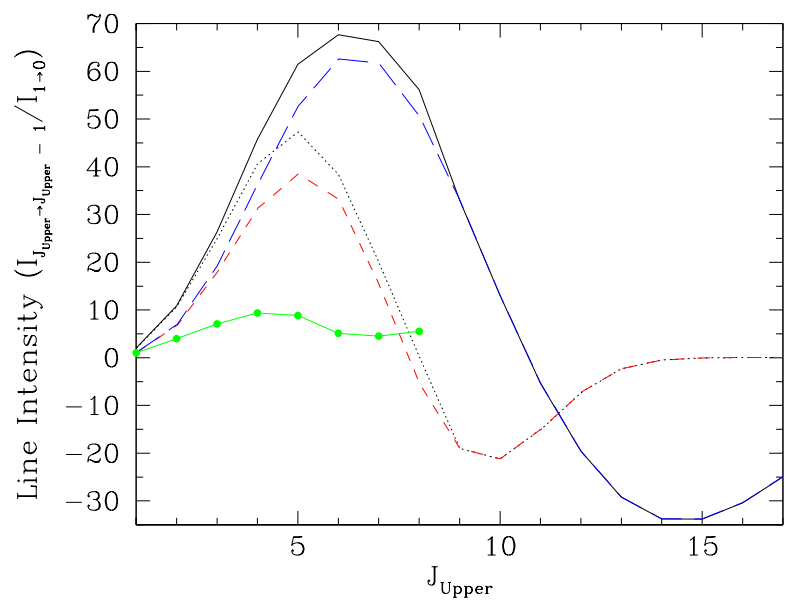}
\caption{Plot of the normalized CO intensity SLED for the Milky Way
  (solid green line; Fixsen et al.\ 1999) and the starburst (blue,
  long-dashed line) and failed (red, short-dashed line) models shown
  in Figure 4. The sum of the Milky Way and starburst SLED yields the
  solid black line, while adding together the Milky Way and the
  failed starburst gives the dotted black line that peaks at lower $J_{\mathrm{Upper}}$ values and has a shallower absorption feature than that of the starburst.}
\end{figure}

\subsection{The Effects of X-ray Dominated Regions}
\label{sect:XDR}
As mentioned above, the current calculations do not include the
effects of X-ray heating by the AGN on the CO SLED. XDRs have been shown in simulations to produce
substantial excitation of CO lines out to very high J values.  The
value of $J_{\mathrm{Upper}}$ at which the flux SLEDs peaks for XDR models can
be considerably greater than 10 (Schleicher et al. 2010), which is not observed for the nuclear
starburst disk models explored here (see Figure 4).  The size of the XDR at the inner radius of the
nuclear starburst disk will depend on the density of the disk, the strength of the X-ray illumination by the
central AGN, and the strength of the starburst in the disk (although
this is a weak dependence). Schleicher
et al. (2010) recently modeled the sizes of XDRs in starbursts near
AGNs assuming typical AGN X-ray luminosities and ISM densities. The
densities of the nuclear starbursts models are very large
($>10^{7}$~cm$^{-3}$ for a failed model and $>10^{9}$ cm$^{-3}$ for
one with a pc-scale starburst), in which case
figure 2 of Schleicher et al. (2010) indicates that the radius of the
XDR would likely be only a fraction of a parsec for a disk with a
pc-scale burst, while the XDR may extend $\sim 1$~pc for the failed
model. If this is accurate then the strong absorption seen in the high
J lines of the pc-scale starburst models (e.g., Figure 4) is robust to the
inclusion of X-ray heating because the absorption is caused by the
material in the outer regions of the starforming disk (Fig. 1). The influence of AGN feedback and its effects
on the calculated CO SLED will be included self-consistently in future work.

\section{Conclusions}
\label{sect:Conclusion}
Nuclear starburst disks are an important potential mechanism to
explain the obscuration of $z < 1$ Seyfert galaxies (Ballantyne 2008). A pc-scale
starburst within the disk can potentially inflate the disk to the
point where it obscures a large fraction of the lines-of-sight to the
AGN and, thus, provides the toroidal obscuration required by the
unified model. At the same time, the disk processes gas and moves it toward the
central engine where it can accrete onto the black hole. The
nuclear starburst disk therefore connects the black hole and galactic
environments, a mechanism necessary to explain the observed
relationships between black holes and their host
galaxies. Observational tests of the presence of these disks are,
therefore, important to verify the validity of the theory and to
characterize the properties of these nuclear starburst disks (Pierce et al. 2011).

Here, we present the predicted CO line emission and absorption properties
of nuclear starburst disks around AGN based on the analytic model of
Ballantyne (2008).  The CO luminosities and flux SLEDs are found to have
quantitative and qualitative features similar to ULIRGs, LIRGs, and
other starburst environments, but with smaller amplitudes due to the
more compact sizes of the starbursts. Direct detection of the nuclear
starburst will require significant sensitivity and resolution to
separate its emission from the surrounding galaxy. Selecting a target
sample of AGN galaxies with weak on-going galactic-scale
star formation (which may be common for $z < 1$ X-ray AGNs; Silverman
et al. 2009) will increase the likelihood of detection. Stacking of
the CO SLEDs from such a sample will be an excellent method to measure
the shape of the SLED out to large $J_{\mathrm{Upper}}$.

Once detected, the CO flux SLEDs contain significant diagnostic
information about the starburst disk. For example, starburst
disks in which a parsec-scale starburst
occurred are found to have larger brightness temperature ratios
between the higher and lower lines, when compared to the ratios of the
models that failed to meet this criterion (see Table 3 and Figure 2).
The finding that an obscuring torus composed of a nuclear starburst
disk should have a characteristic CO SLED shape is of particular
interest, as it could provide an observational test to discern the
structure of the nuclear regions of active galaxies. In addition, the
large densities and temperatures, as well as steep radial gradients of these variables, associated with the pc-scale starburst produce a CO
SLED that drops into absorption for $J_{\mathrm{Upper}}>10$. This
prediction is robust to the presence of a XDR for typical
Seyfert X-ray luminosities and would be a clear indication of the
ultra-compact nature of the nuclear starburst.

\acknowledgments
This work was supported in part by NSF award AST 1008067 to DRB. The
authors thank D.\ Narayanan and N.\ Murray for useful discussions, and
the anonymous referee for very useful comments that improved the
paper.

\end{document}